\documentclass{mn2e}
\begin{document}
\input epsf
\title[Variable stars in NGC 3201] 
{Variable stars in the globular cluster NGC 3201. I. Multimode 
SX~Phe-type variables}
\author[B. Mazur, W. Krzemi\'nski, I.B. Thompson] 
{Beata~Mazur,$^1$\thanks{E-mail: batka@camk.edu.pl(BM); wojtek@roses.lco.cl(WK);
ian@ociw.edu (IBT)}
Wojciech~Krzemi\'nski$^{1,2}$,
and Ian B. Thompson$^3$\\
$^1$Copernicus Astronomical Center, Bartycka 18, 00-716 Warsaw, Poland\\
$^2$The Observatories of the Carnegie Institution of Washington, Las
Campanas Observatory, Casilla 601, La Serena, Chile\\
$^3$The Observatories of the Carnegie Institution of Washington, 
813 Santa Barbara St., Pasadena, CA 91101
}
\maketitle

\begin{abstract} 
We report on the discovery of eleven multimode SX~Phoenicis--type blue
stragglers in the field of the southern globular cluster NGC~3201.  In
these variables both radial and non-radial modes are excited.  For
three variables the derived period ratio is close to that observed in
SX~Phoenicis itself, suggesting that these stars are pulsating in the
fundamental and the first-overtone radial modes.  Using the McNamara
(1997) period-luminosity relation we have estimated the apparent
distance modulus to NGC~3201 to be 14.08$\pm0.06\pm0.1$mag.
\end{abstract}


\section  {Introduction}
Variable stars in star clusters have long been recognized as unique
laboratories that allow the study of both the properties of the
variables and the stellar systems they belong to.  One of the most
important applications is the use of the variables to determine the
distance to the parent system.  Interest in SX~Phoenicis variables
as precise distance indicators has grown in the last few years with
the publication of results of the {\it Hipparcos} mission (ESA 1997).
McNamara (1997) used a semiempirical approach to determine the
period-luminosity (P-L) relation for high-amplitude $\delta$ Scuti (HADS)
stars; the zero point of this relation was set with the aid of {\it
Hipparcos} parallaxes.  Petersen \& H\o g (1998) derived a P-L relation
for HADS which was based on {\it Hipparcos} parallaxes of six
variables, including SX~Phoenicis itself.  Both relations yield
distances with an uncertainty of $\pm 0.10 $ mag.  Theoretical
calculations imply that the inclusion of colour and metallicity terms
can significantly decrease the intrinsic scatter of this relation
(Petersen \& Christensen-Dalsgaard 1999, Santolamazza et al. 2001).
Thus SX~Phoenicis stars can provide a valuable cross-check for
the distance determined with  other methods, e.g. with  other
types of variable stars: detached eclipsing binaries (Paczynski 1997),
W~UMa-type systems (Rucinski \& Duerbeck 1997) and RR~Lyraes.

In 1989 we started at Las Campanas Observatory a CCD survey for
short-period variables in southern star clusters. The sample contained
open clusters (e.g. Cr261 -- Mazur, Krzemi\'nski \& Kaluzny 1995,
NGC~2243 -- Kaluzny, Krzemi\'nski \& Mazur 1996, Berkeley~39 -- Mazur,
Krzemi\'nski \& Kaluzny 1999a) and globular clusters (e.g. NGC~6362 --
Mazur, Kaluzny \& Krzemi\'nski 1999b and references therein).  The
results of the survey have shown that CCD photometry can be an
effective method of detection of pulsating stars and short-period
binaries, even those with periods of a few days. However an extended
data set is a must to detect such binaries.  In a continuation of this
survey we embarked in January 1993 on an extended CCD study of the
globular cluster NGC~3201. Our goal was to obtain as complete a  census
of variable stars as possible, with the aim of using observations of the
variables to derive the distance modulus to the cluster.  

NGC~3201 ($\alpha_{2000}=10^{\rm h}17^{\rm m}37^{\rm s},
\delta_{2000}=-46\degr 24\fm 7, l=277\fdg 2, b=8\fdg 6$), is a nearby
cluster of low central concentration, and has been the subject of
numerous photometric and spectroscopic studies.  Historically, the
first papers were devoted to variable stars (Woods 1919, Bailey 1922,
Dowse 1940, Wright 1941, Wilkens 1965, Cacciari 1984, Samus et al. 1996).
Photographic
CMDs were published by Alcaino (1976), Lee (1977) and Alcaino \& Liller
(1981), and CMD's based on  CCD photometry by Penny (1984), Alcaino,
Liller \& Alvarado (1989), Brewer et al. (1993), Covino \& Ortolani
(1997), Rosenberg et al. (2000), and von Braun \& Mateo (2002).  Cote
at al. (1994) presented radial velocities of a sample of the cluster
giants, they also used these multiple velocity measurements to search
for binaries.

The metal abundance of NGC~3201 falls in the intermediate range,
however the exact value is still not well established.  Zinn \& West
(1984) derived  [Fe/H]=--1.61 and Carretta \& Gratton (1997) obtained
[Fe/H]=--1.23 from an analysis of echelle spectra of 3 cluster giants.
Gonzalez \& Wallerstein (1998) found a mean value of [Fe/H]=--1.42 from
the echelle spectra of 18 giants.

The Catalogue of Variable Stars in Globular Clusters (Clement et al.
2001) lists 100 entries for NGC~3201, 8 of them with the notation ''not
var''.  Of the 92 variables, 77 are  RR~Lyrae stars.  Recently, von
Braun \& Mateo (2002), reported  the discovery of 14 short-period
variables, mainly eclipsing binaries, in the cluster field.  Using
radial velocities and  location of the variables in the cluster CMD
they concluded  that only one of their variables, a W~UMa-type blue
straggler, is a cluster member.  No SX~Phoenicis-type variables has
been reported thus far.

\section {Observations and data reductions} Observations reported in
this paper were obtained with the 1 m Swope and 2.5 m duPont
telescopes of the Las Campanas Observatory with a variety of CCD
cameras. A field centered on the cluster core was observed during 24
nights between 1993 Jan 6 and 1994 Jan 3, {\sc{ut}}.
The log of the observations is summarized in Table 1. 

\setcounter{table}{0}
\begin{table}
\begin{minipage}{80mm}
\caption{Journal of the observations.}
\label{tab1}
\begin{tabular}{@{}r c c c c c}
Date UT     & Nobs  &$\Delta t$ & Tel.      &  Field of view\\
            &  $V$  &h~m    &               &  arcmin$^2$    \\
            &       &       &               &               \\
Jan ~6,  93 &   ~42 & 4:25  &   1 m          & ${ 5.8\times 5.8}$ \\
Jan ~7,  93 &   ~44 & 4:26  &   1 m          & ${ 5.8\times 5.8}$ \\
Jan ~8,  93 &   ~26 & 2:23  &   1 m          & ${ 5.8\times 5.8}$ \\
Jan 11,  93 &   ~48 & 4:43  &   1 m          & ${ 5.8\times 5.8}$ \\
Feb ~2,  93 &   ~68 & 7:18  &   1 m          & ${12\times 12}$ \\
Feb ~3,  93 &   ~76 & 7:30  &   1 m          & ${12\times 12}$ \\
Feb ~4,  93 &   ~72 & 7:20  &   1 m          & ${12\times 12}$ \\
Mar 12,  93 &   ~78 & 7:42  &   1 m          & ${12\times 12}$ \\
Mar 13,  93 &   ~47 & 4:34  &   1 m          & ${12\times 12}$ \\
Dec 18,  93 &   ~19 & 2:09  &   1 m          &${ 12\times 12}$ \\
Dec 19,  93 &   ~29 & 3:31  &   1 m          &${ 12\times 12}$ \\
Dec 20,  93 &   ~29 & 3:18  &   1 m          &${ 12\times 12}$ \\
Dec 21,  93 &   ~31 & 3:52  &   1 m          &${ 12\times 12}$ \\
Dec 22,  93 &   ~36 & 3:56  &   1 m          &${ 12\times 12}$ \\
Dec 23,  93 &   ~36 & 3:56  &   1 m          &${ 12\times 12}$ \\
Dec 24,  93 &   ~36 & 3:45  &   1 m          &${ 12\times 12}$ \\
Dec 27,  93 &   ~22 & 2:48  &   1 m          &${ 10.5\times 10.5}$ \\
Dec 28,  93 &   ~25 & 2:56  &   1 m          &${ 10.5\times 10.5}$ \\
Dec 29,  93 &   ~23 & 2:38  &   1 m          &${ 10.5\times 10.5}$ \\
Dec 30,  93 &   ~18 & 2:06  &   1 m          &${ 10.5\times 10.5}$ \\
Dec 31,  93 &   ~24 & 2:37  &   1 m          &${ 10.5\times 10.5}$ \\
Jan 01,  94 &   ~66 & 3:56  &   2.5 m        &${ 4.4\times 4.4}$ \\
Jan 02,  94 &   ~54 & 3:54  &   2.5 m        &${ 4.4\times 4.4}$ \\
Jan 03,  94 &   ~50 & 2:40  &   2.5 m        &${ 4.4\times 4.4}$ \\
\end{tabular}
\end{minipage}
\end{table}

In January 1993 the Texas Instrument TI\#1 $800 \times 800$ CCD was
used, giving a field size of 5.8 arcmin with a scale of 0.435
arcsec/pixel. In February and March 1993 data were taken with the
Tektronix \#1 $1024 \times 1024$ CCD with a scale of 0.7 arcsec/pixel,
and the field size of 12 arcmin.  The same CCD was used at the
beginning of the December 1993 run, later it was replaced with the
Tektronix \#2 $1024 \times 1024$ chip, with a scale of 0.61
arcsec/pixel and a field size of 10.5 arcmin.  The Tek\#1 camera was used
again during the January 1994 run on the duPont telescope, this time
the field of view was 4.4 arcmin with a scale of 0.21 arcsec/pixel.
Most of the frames were collected in the $V$ band, in addition several
frames were taken with $B$ and $I$ filters.  The exposure time for the
$V$ frames taken with the Swope telescope was 300-360 s, whereas
for those collected with duPont telescope it was 120 s. We obtained a
total of 999 frames in the $V$ band.  On the night of 1994 January 2
several standard stars from Landolt (1992) were observed.  They were
used to determine the transformations to the standard $BV$ system.

The preliminary processing of the raw CCD data was performed under {\sc
iraf} \footnote{{\sc iraf} is distributed by National Optical Astronomy
Observatories, which is operated by the Association of Universities for
Research in Astronomy, Inc., under cooperative agreement with the
National Science Foundation.}.  All frames were bias subtracted and then
flatfielded with median-combined sky flats.  Photometry of NGC 3201
stars was measured using two software packages: frames used to extract
the cluster CMD were analyzed with {\sc daophot} (we used a version
with variable PSF; Stetson 1987, 1989), while the sequences of
observations obtained for searches for variables were reduced with {\sc
d{\footnotesize o}phot} package (Schechter, Mateo \& Saha 1993). {\sc
d{\small o}phot V1.1} uses an analytic, position-independent point
spread function (PSF). In order to minimize the influence of variable
PSF on the light curves of individual stars we applied a method of
local comparison stars, as described in detail in Kaluzny et al.
(1993).
 
\section {Variables}
Sequences of observations in the $V$ filter were used to search for the
variables in the cluster field.  Candidate variables were
identified following two methods described in detail in Mazur,
Krzemi\'nski \& Kaluzny (1995). First, we applied the traditional
``scatter search'' technique -- we selected stars exhibiting excessive
scatter compared to other stars of comparable brightness. The idea for
the second method was taken from Stellingwerf's (1978) paper on the
phase dispersion minimization (PDM) method. We divided the overall
variance calculated for consecutive intervals of a time-domain light
curve by the global variance obtained from all data points and
selected for further examination stars for which the quotient
was the smallest.  An additional examination was made of light
curves of all stars that fall in the blue straggler region.

Analysis of the 1993/94 data set led to the identification of 32 new
variable stars (Mazur 1996), half of which are SX~Phe--type pulsating
blue stragglers. In eleven of these, multiple modes  are excited and
these stars will be discussed in the present paper.

In Fig. 1 we present finding charts for the newly found multimode SX~Phe-type variables. 

\begin{figure*}
\epsfxsize=15.2cm\epsffile{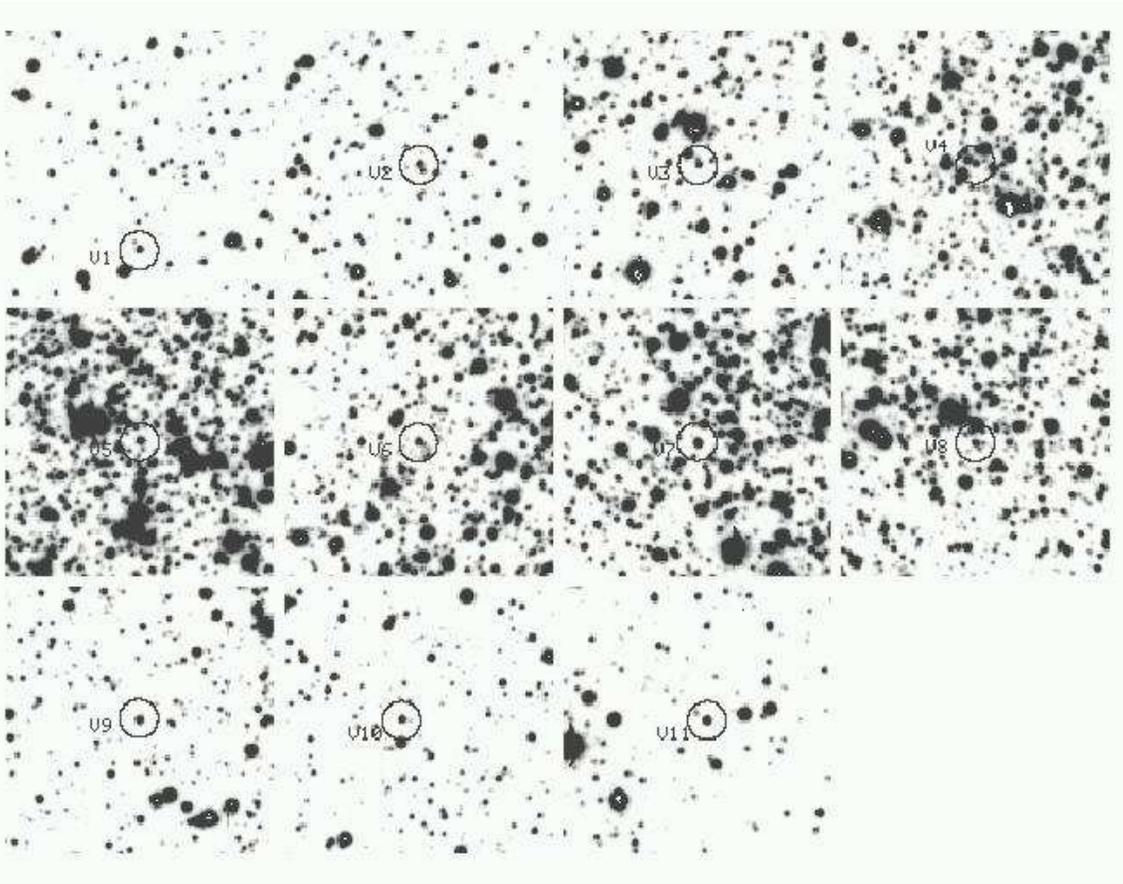}
  \caption{The $V$-band finder charts for the 11 newly discovered multimode SX~Phe-type variables.
Each chart is 60 arcsec on a side with north up and east to the left. }
  \label{chart}
\end{figure*}

\setcounter{table}{1}
\begin{table}
\begin{minipage}{80mm}
\caption{Coordinates of NGC~3201 multi-mode SX Phe type variables.}
\label{tab2}
\begin{tabular}{@{}l r r r r }
var&$\alpha $(2000)~ & $\delta $(2000)~& X $\arcsec$  &  Y $\arcsec$\\
    &{\rm h~~m~~s~~~}   &\degr~ $\arcmin~~\arcsec$&           &          \\           
V1 &     10 17 43.5 & -46 21 53.4 & 67    & 171     \\
V2 &     10 17 32.1 & -46 22 37.9 & -53   & 126     \\
V3 &     10 17 36.6 & -46 23 \phantom{0}3.4& -5  & 100\\
V4 &     10 17 42.6 & -46 24 19.5 &  58    & 24       \\
V5 &     10 17 36.0 & -46 24 38.2 & -11    & 4        \\
V6 &     10 17 42.5 & -46 24 49.1 & 57     & -7       \\
V7 &     10 17 40.2 & -46 25 \phantom{0}7.8 & 32   & -26       \\
V8 &     10 17 36.0 & -46 25 40.2 & -11   & -59       \\
V9 &     10 17 48.1 & -46 25 53.4 & 115   & -78       \\
V10&     10 17 50.9 & -46 25 57.7 & 145   & -76       \\
V11 &     10 17 20.1 & -46 26 38.0 & -178  & -118       \\

\end{tabular}
\end{minipage}
\end{table}

Table 2 lists the coordinates of the variables.  Column (1) gives the
designation of the star; column (2) and (3) are the equatorial
coordinates, derived from the position of about 100 stars
selected from the USNO-A V1.0 Catalogue of Astrometric
Standards (Monet at al. 1996). The errors of the  coordinates are about
1 arcsec.  Columns (4) and (5) are the coordinates in the system of the
Sawyer Hogg Catalogue (Sawyer Hogg 1973).
 
Fig. 2 shows the locations of the variables in the cluster
colour-magnitude diagram (CMD).  The $V/B-V$ diagram is based on a
100~s $V$-band exposure  and a 200~s $B$-band exposure taken with the
duPont telescope on the night of 1997 February~20~{\sc ut}.  The field
was centered on the cluster core and covered an area of $8.8\times 8.8$
arcmin.  The CMD is not intended to be complete -- stars having
atypically large photometric errors at a given  magnitude were
omitted.  All of the variables are located in the blue straggler
region, and occupy a narrow strip of width 0.11 mag in $B-V$.

\begin{figure}
  \epsfxsize=7.8cm\epsffile{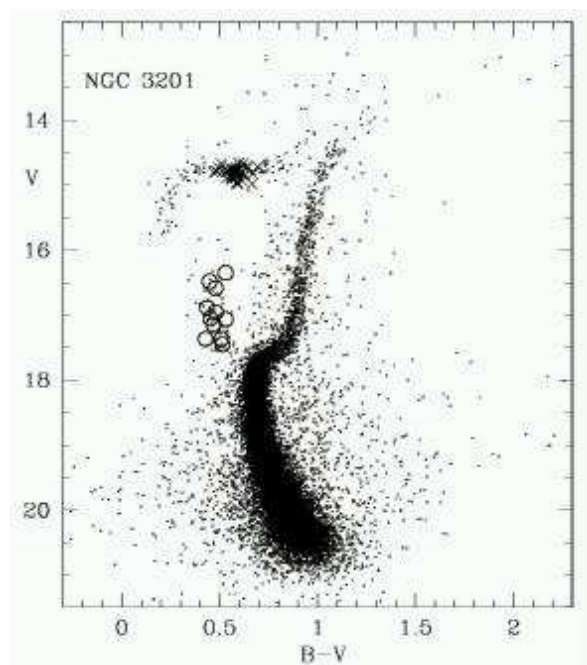}
  \caption{The $V/B-V$ CMD for NGC~3201 (see text for details).
   Open circles mark positions of SX~Phoenicis stars, corresponding
   to mean colours and magnitudes. Crosses are RR~Lyrae stars. }

   \label{fig2}
\end{figure}
	
Table 3 lists the photometric properties of the variables.   

\setcounter{table}{2}
\begin{table}
\begin{minipage}{80mm}
\caption{Photometric properties of variables.}
\label{tab3}
\begin{tabular}{@{}l r r l  l l l l l l l }
 var&$<V>$     &$B-V$ & Frequency(1/d) &${A}_{V}$& mode \\
    &            &      &         &      &      \\   
V1  &    17.15   & 0.47 & 22.057 & 0.05   &       \\
    &            &      & 22.380 & 0.025  &        \\
    &            &      &         &      &      \\   
V2  &    17.37   & 0.51 & 26.842 & 0.04   &  F  \\
    &            &      & 34.350 & 0.05   &  1OT      \\  
    &            &      &         &      &      \\   
V3  &    17.46   & 0.51 & 26.634(27.634) &  0.025  &    \\
    &            &      & 27.968(28.967) &  0.025  &     \\
    &            &      &         &      &      \\   
V4  &    17.37   & 0.43 & 26.567 & 0.07   &    \\
    &            &      & 27.020 & 0.04   &       \\
    &            &      &         &      &      \\   
V5  &    16.89   & 0.44 & 22.991(21.991) & 0.04   &    \\ 
    &            &      & 29.367(30.368) & 0.025  &    \\
    &            &      & 28.204(29.204) & 0.02   &    \\
    &            &      &         &      &      \\   
V6  &    16.96   & 0.48 & 24.7035 & 0.13 & 1OT    \\
    &            &      & 25.252 & 0.07 &          \\
    &            &      &         &      &      \\   
V7  &    16.35   & 0.53 & 14.85881 & 0.37 & F   \\
    &            &      & 19.06239 & 0.22 & 1OT        \\
    &            &      &         &      &      \\
V8  &    17.05   & 0.45 & 18.42251 & 0.39 & F      \\
    &            &      & 23.619  & 0.03 &  1OT   \\
    &            &      &         &      &      \\
V9  &    16.59   & 0.48 &  19.92607 & 0.36 &  1OT  \\
    &            &      &  30.468  & 0.025 &   \\  
    &            &      &  19.047(20.047) & 0.023  \\
    &            &      &         &       &      \\
V10 &    17.07   & 0.53 &  20.874 & 0.03  &   \\
    &            &      &  21.301 & 0.025 &     \\
    &            &      &         &      &      \\   
V11 &    16.48   & 0.45 & 18.46393 & 0.2  & 1OT  \\
    &            &      & 18.273  & 0.03 &     \\
    &            &      & 18.677  & 0.03 &   \\
\end{tabular}

\end{minipage}
\end{table}

Column (2) gives mean $V$-band  magnitude.
Column (3) is the $<B>-<V>$ colour, while column (4) shows the frequencies
of pulsations, in cycles per day. 
Column (5) gives the full amplitude in the $V$ band.
The mean $<B>$ magnitudes were derived from data obtained in February 1997 on the duPont telescope.

The data have been collected during 6 observing runs which separate
into two data sets. The first data set consist of January, February and
March 1993 runs and we will refer to it as to season 1993; the second
set is formed by 3 runs in December 1993/January 1994 -- hereafter
season 1994. Since the gap between both data sets is long (9 months),
we calculated Fourier periodograms for each season separately. The long
gaps between successive runs and the fact that these runs were very
short resulted in a very complex spectral window for the 1993 season.
In the case of low amplitude variables, the noise introduced by
photometric errors made it impossible  to decide which of the aliases
is real.  The spectral window for the 1994 season is much clearer
although the resolution is poorer.
We decided to restrict our analysis to the 1994 data. We used the 1993 data set
to improve the determination of frequencies of high amplitude modes.

Fourier periodograms of the 11 SX~Phe--type pulsators
for the 1994 season are shown in Figures~3-13.  In most cases there is
a considerable signal at very low frequencies. Most likely the presence
of these frequencies is caused by zero-point drift, which in turn seems
to be a result of variations of the PSF across the chip.  The method of
local comparisons can compensate for errors in the profile photometry
to some extent, but some variability on a time-scale of nights
remains.  An additional source of error is the combination of
photometry that was collected with 3 different telescope/chip
configurations.  We assume  that the low frequencies present in the
periodograms are not intrinsic to the variables, and decided to fit a
sine wave to the data and subtract this long term variability;
frequencies less than 0.2 cycle {${\rm d^{-1}}$ have been removed from the data.
\begin{figure}
\epsfxsize=8.3cm \epsffile{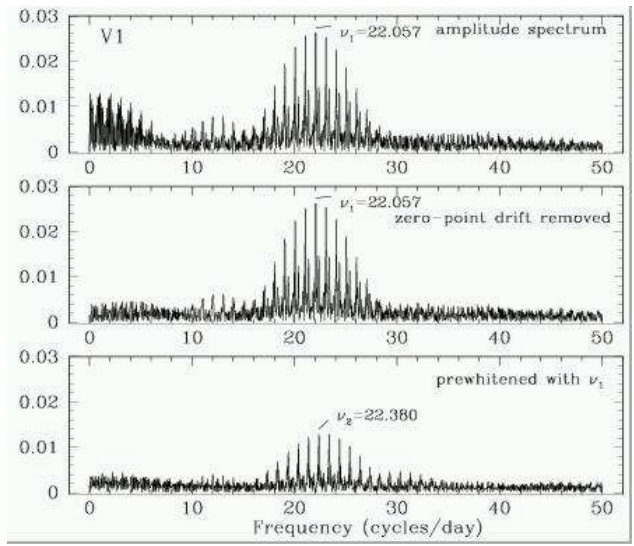}
  \caption{The amplitude spectra of SX~Phe star V1.
  Panels show Fourier transform of all data, after removal of zero-point drift,
  prewhitened with dominant mode, respectively.}
\label{v1}
\end{figure}
       
\begin{figure}
\epsfxsize=8.3cm \epsffile{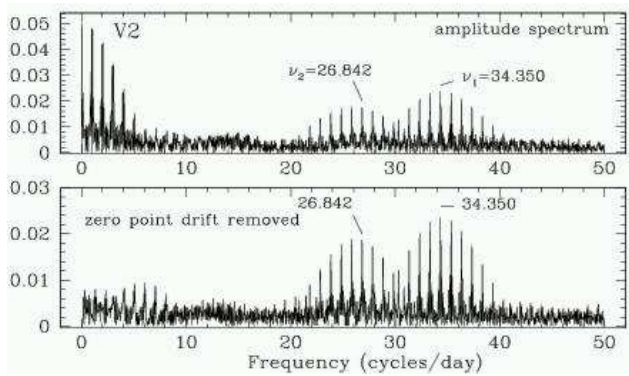}
 \caption{The amplitude spectra of SX~Phe star V2.
  Panels show Fourier transform of all data and after removal of zero-point drift,
  respectively.}
\label{v2}
\end{figure}

\begin{figure}
\epsfxsize=8.3cm \epsffile{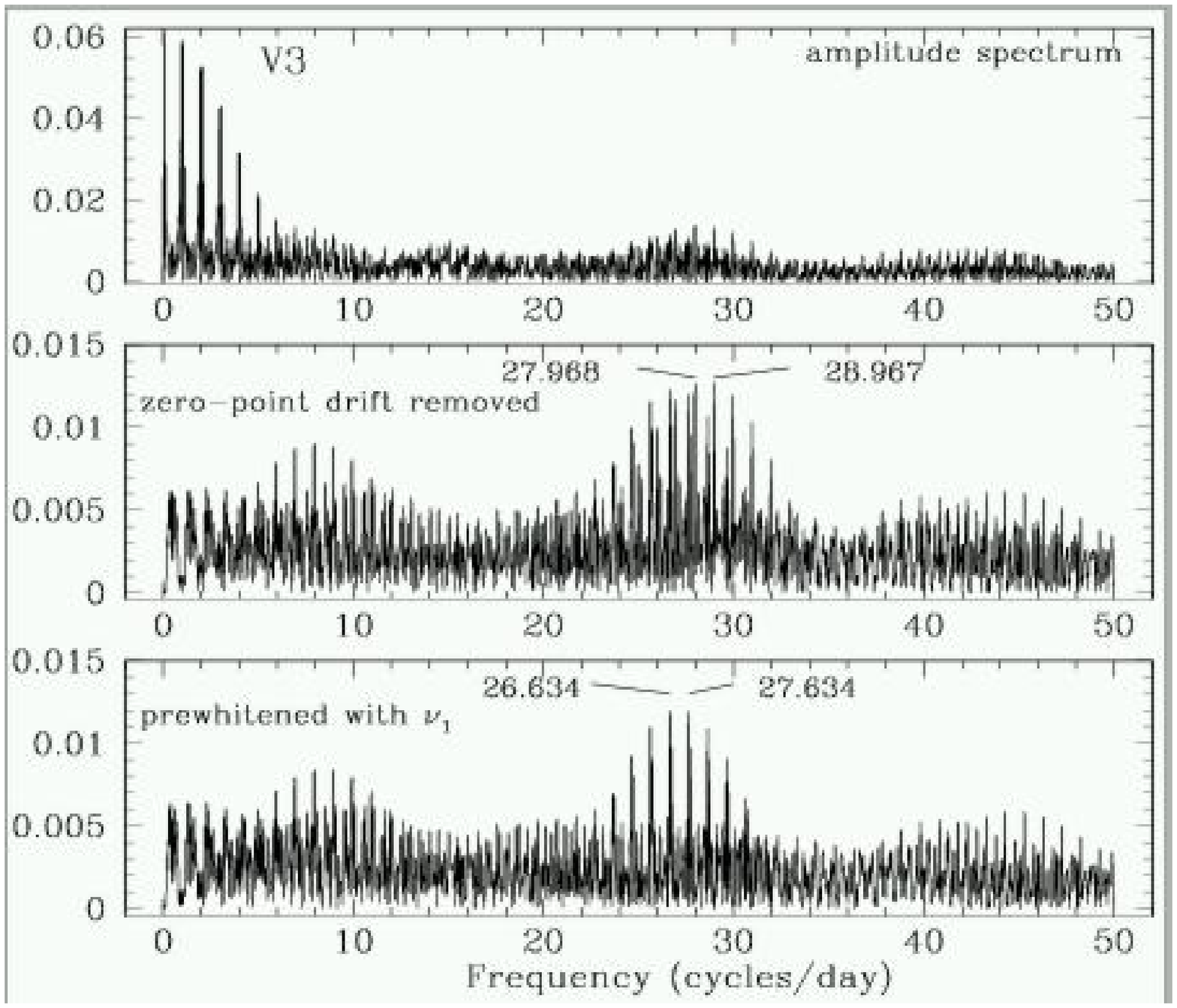}
\caption{ The amplitude spectra of SX~Phe star V3.
Panels show Fourier transform of all data, after removal of zero-point drift,
prewhitened with dominant mode, respectively.}
\label{v3}
\end{figure}
 
\begin{figure}
\epsfxsize=8.3cm \epsffile{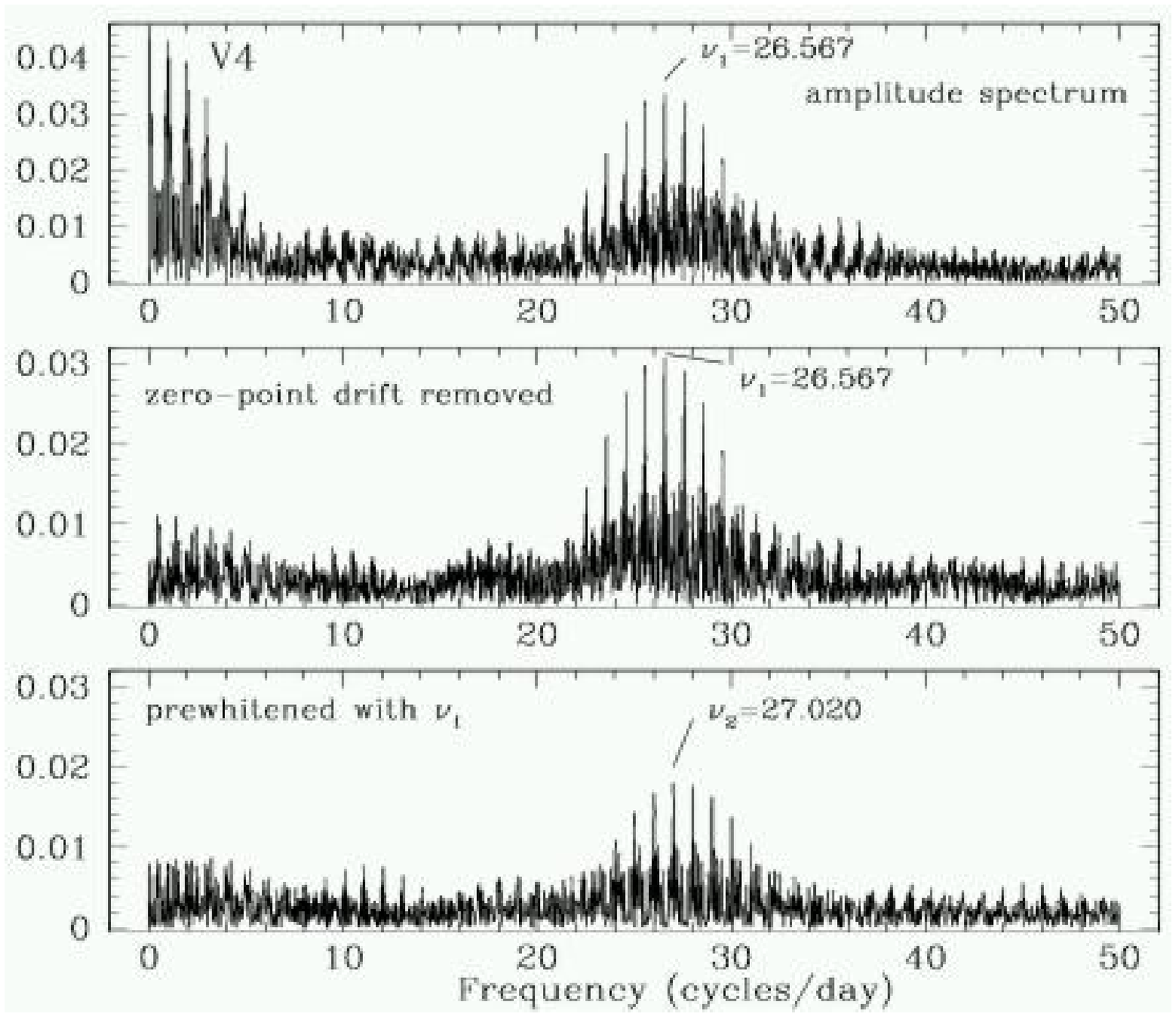}
\caption{The amplitude spectra of SX~Phe star V4.
Panels show Fourier transform of all data, after removal of zero-point drift,
prewhitened with dominant mode, respectively.}
\label{v4}
\end{figure}
\begin{figure}
\epsfxsize=8.3cm \epsffile{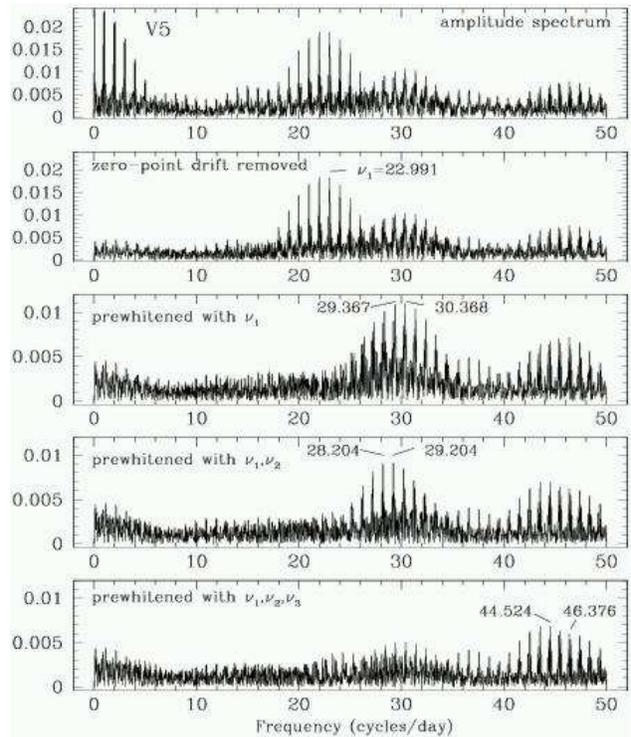}
\caption{The amplitude spectra of SX~Phe star V5.
Panels show Fourier transform of all data, after removal of zero-point drift,
prewhitened with the dominant, the second and the third modes, respectively.}
\label{v5}
\end{figure}

Fourier transforms have been recalculated, and the frequency of the
highest peak has been used as an input to the programme NFIT
(Schwarzenberg-Czerny 1999). NFIT fits a Fourier series by nonlinear
least squares allowing for a refinement of frequency; several base
frequencies, their harmonics and linear combinations can be included.
Once the first frequency was determined, the data were prewhitened and
the Fourier transform of the residuals was calculated. The second main
frequency was identified and NFIT was used to fit a Fourier series of
both frequencies to the data. The number of harmonics and the
combination terms depended on the amplitude of the excited modes. In
the case of the low amplitude variables for each frequency usually only
one term was significant. In the case of the high amplitude double-mode
variable V7  combination terms up to fourth order could be detected.
The second solution was again subtracted from the data and the
residuals were searched for additional periodicities.  For eight
variables the Fourier transform of the residuals from the two frequency
solution showed no more periodicities with the full amplitudes in the
$V$-band in excess of 0.02 mag.  In the case of V5, V9 and V11 a third
periodicity has been detected, and the whole procedure was repeated.
In some cases the Fourier transform showed daily aliases of comparable
height. This problem could generally be resolved  with the help of the
1993 data when the observing nights were longer and the side-lobes due
to day cycles smaller. We have also compared residuals resulting from fits
assuming different combinations of frequencies, but in the case of low amplitude
variables differences we got were negligible. 
In several cases we were not able to distinguish between 
a real frequency and an alias, and two possibilities are given in Table 3.

\begin{figure*}
\epsfxsize=14.5cm \epsffile{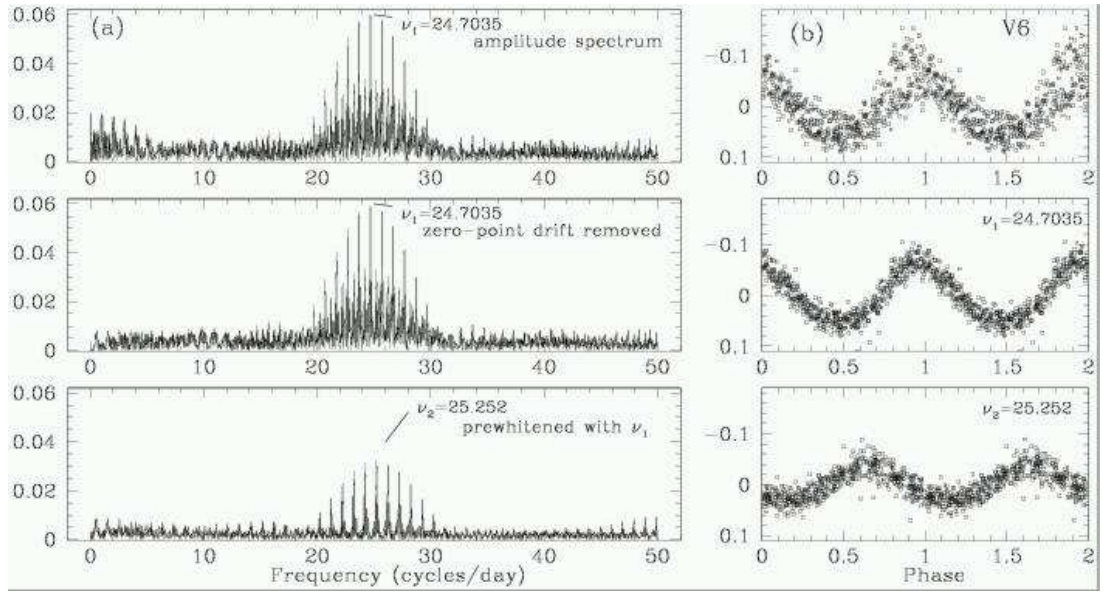}
\caption{(a) The amplitude spectra of SX~Phe star V6.
Panels show Fourier transform of all data, after removal of zero-point drift, 
prewhitened with dominant mode, respectively.
(b) Light curves for V6. The uppermost panel
shows data after removal of the zero-point drift.
Lower panels show light curves
obtained by subtracting the variability corresponding to one of two periods,
its harmonics and combination terms.}
\label{v6}
\end{figure*}

\begin{figure*}
\epsfxsize=14.5cm \epsffile{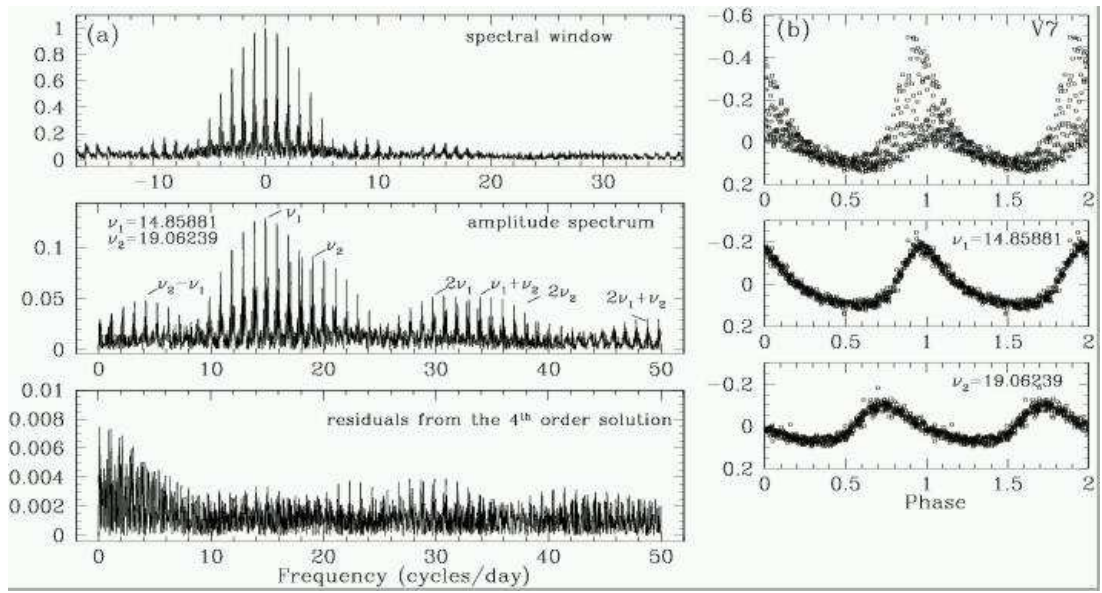}
\caption{(a) The spectral window and the amplitude spectrum of SX~Phe star V7.
(b) Light curves for V7. The uppermost panel
shows data phased with the period of fundamental mode.
Lower panels show light curves
obtained by subtracting the variability corresponding to one of two periods,
its harmonics and combination terms.}
\label{v7}
\end{figure*}

\section{Mode identification}
Multiperiodicity of SX~Phe variables opens the possibility of applying
the theory of pulsations to constrain masses and possible formation
scenarios of blue stragglers (Gilliland \& Brown 1992, Petersen \&
Christensen-Dalsgaard 1996).  The first step in constructing  models is
the identification of modes.  Until recently,  SX~Phe variables were
usually attributed a single (fundamental or first-overtone) radial mode
(Nemec \& Mateo 1990; Nemec, Linnell Nemec \& Lutz 1994, McNamara
1995).  SX~Phe itself was for a long time the only member of the class
known to pulsate in fundamental and first-overtone modes
simultaneously.  This situation has changed in the last few years.
Garrido \& Rodriguez (1996) analysed the  existing data on SX~Phoenicis
to show that in addition to two high-amplitude modes at least two other
periodicities are present (one of them clearly non-radial) with
amplitudes at the mmag level.  Additional periodicities have also been
discovered in another field SX~Phe-type star, BL Camelopardalis (Hintz
et al.1997, Zhou et al. 1999).  Gilliland et al. (1998) reported on the
discovery of multimode SX~Phoenicis pulsators in 47~Tucanae. For one of
the variables (V2) the interpretation of the detected periodicities as
fundamental radial mode and first overtone is based on the derived
period ratio (0.772).  For the others, a safe identification of modes
is not possible (Bruntt et al. 2001).  Pych et al. (2001) discovered
twelve double-mode variables among SX~Phe stars in M55. For the
majority of these the  period ratio is close to 1, implying that at
least one of the excited modes is nonradial.  Freyhammer, Petersen \&
Andersen (2000) reported the discovery of 11 multimode pulsators in
$\omega $ Cen, and they estimate that $\omega $ Cen hosts hundreds of
SX~Phe stars.  Petersen et al. (2000) analyzed the published light
curves of 48 SX~Phe variables in globular clusters in search for
double-mode oscillation.  Several stars that show indications for
double-mode pulsations in low order radial modes have been identified.

\begin{figure}
\epsfxsize=8.3cm \epsffile{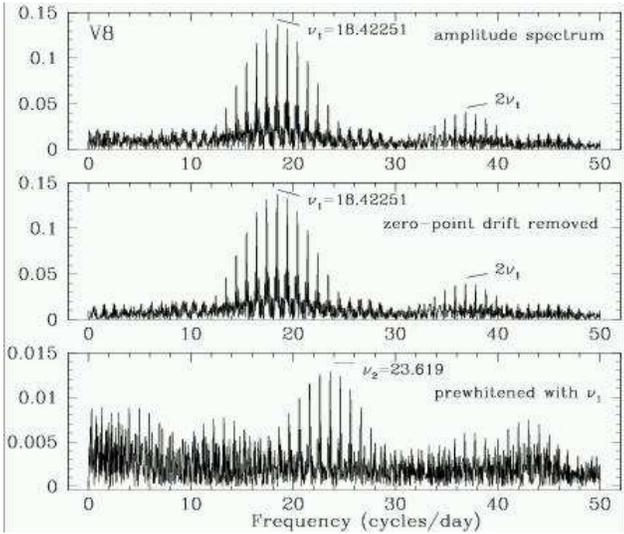}
\caption{The amplitude spectra of SX~Phe star V8. 
Panels show Fourier transform of all data, after removal of zero-point drift,
prewhitened with dominant mode, respectively.}
\label{v8}
\end{figure}

\begin{figure}
\epsfxsize=8.3cm\epsffile{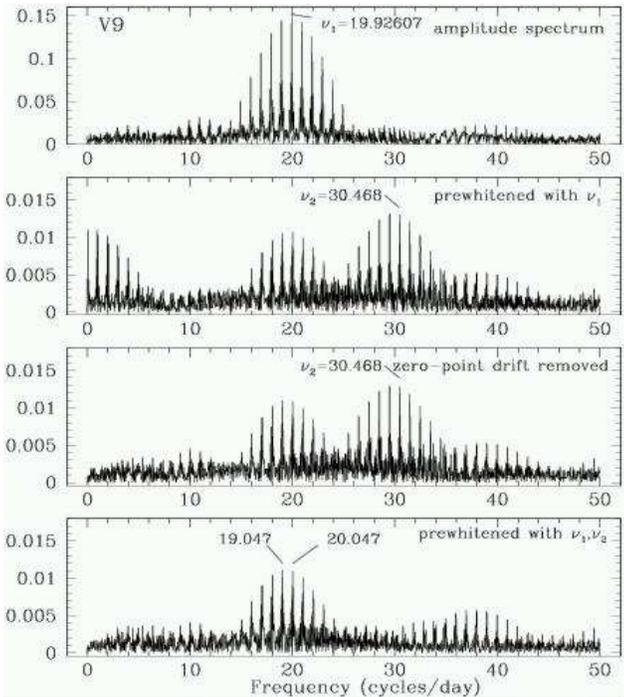}
  \caption{The amplitude spectra of SX~Phe star V9.
Panels show Fourier transform of all data, after removal of zero-point drift,
prewhitened with the dominant and the second  modes, respectively.}
    \label{v9}
    \end{figure}

\begin{figure}
\epsfxsize=8.3cm \epsffile{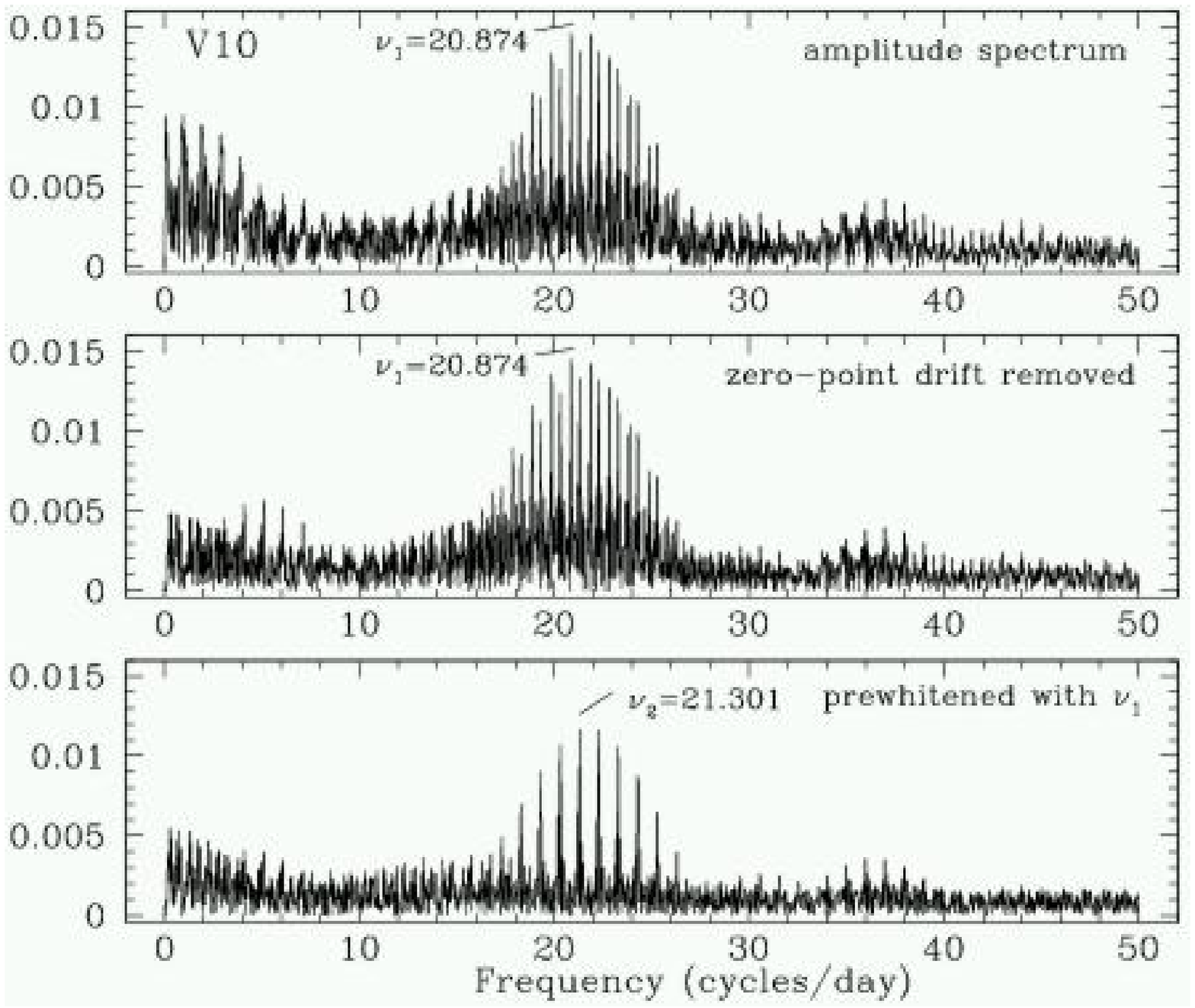}
\caption{ The amplitude spectra of SX~Phe star V10.
Panels show Fourier transform of all data, after removal of zero-point drift,
prewhitened with dominant mode, respectively.}
\label{v10}
\end{figure}

\begin{figure}
\epsfxsize=8.3cm\epsffile{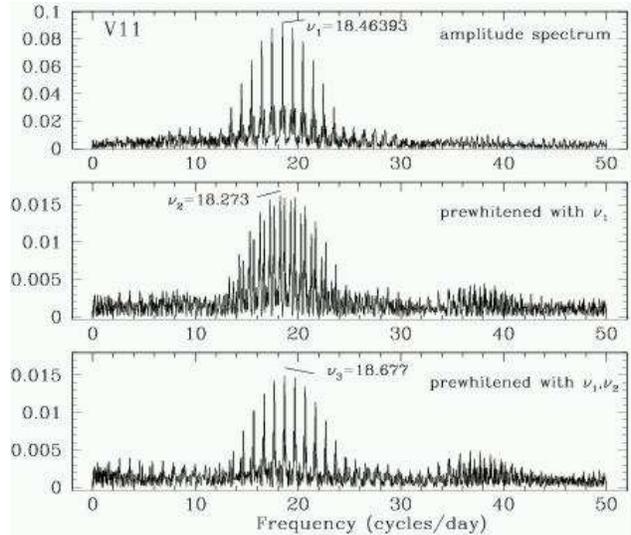}
  \caption{ The amplitude spectra of SX~Phe star V11.
  Panels show Fourier transform of all data,
  prewhitened with the dominant and the second  modes, respectively.}
  \label{v11}
\end{figure}

All these studies show that in SX Phoenicis stars both radial and
nonradial modes are excited suggesting that they are  population II
twins of $\delta$ Scuti variables.  Multisite photometric campaigns and
spectroscopic studies suggest that in $\delta$ Scuti stars hundreds of
nonradial pulsation modes are simultaneously excited (Breger 2000).
Only the low-degree modes can be detected photometrically. The
multisite observing campaigns resulted in the discovery of about 20
independent modes in the best studied stars (FG Vir - Breger et al.
1998, 4 CVn - Breger et al. 1999).  The number of the observed
frequencies in SX~Phe-type stars is much smaller, and this is most
likely due to the lower accuracy of the data and the shorter timespan
of observing runs -- many of the peaks in $\delta$ Scuti stars fall in
the mmag range.

The photometric methods that can be employed to discriminate between
radial and low-order nonradial modes rely on ``amplitude ratios versus
phase shifts'' diagrams. In order to apply this analysis technique
multicolor photometry is required [$BV$ -- Watson (1988), Str\"omgren
$vby$
 -- Garrido, Garcia-Lobo \& Rodriguez (1990].  In the 1993/1994 seasons
we observed NGC~3201 in mainly the $V$ band, the few obtained frames in
$B$ do not allow mode discrimination.  However, there is an indirect
method that one can use. Rodriguez et al. (1996) examined the amplitude
ratios and phase shifts between observed light and color variations to
show that the high amplitude (with full visual amplitude greater than
roughly 0.3 mag) $\delta$ Scuti and SX~Phoenicis stars pulsate in
radial modes.  They also analysed 3 monoperiodic medium amplitude
$(0.09<A_V<0.2)$ stars and concluded that they are radial pulsators as
well. Hence we will assume that the high amplitude modes in V7,V8 and
V9 correspond to radial pulsation. It seems likely that the medium
amplitude modes in V6 and V11 are radial as well.  
However this latter assumption may not be correct.
As pointed out by the referee some $\delta$ Scuti stars have nonradial
modes with individual amplitudes in excess of 0.1 mag
(e.g. 1 Mon,  Balona et al. 2001).

Another method that can be useful in  mode identification is the
analysis of observed period ratios.  Petersen \& Christensen-Dalsgaard
(1996) have discussed the $P_1/P_0$ (first overtone to the fundamental
mode) and $P_2/P_1$ (second to first overtone) period ratios in terms
of metal content and mass-luminosity relation.  The calculated period
ratios show a significant dependence on both metal abundance and the
$M-L$ relation. For Z=0.001 they obtain $P_1/P_0$ in the range
0.775-0.790, depending on the $M-L$ relation used.  The uncertainty of
period ratios calculated from models is large but fortunately there is
the possibility of comparing them with the observed period ratio of SX
Phoenicis itself.  The metallicity of SX Phoenicis ([Fe/H]=-1.37 --
Hintz, Joner \& Hintz 1998) is very similar to that of NGC 3201
([Fe/H]=-1.42, Gonzalez \& Wallerstein 1998).  The $P_1/P_0$ period
ratio observed in SX Phoenicis is 0.7782, very close to the value
derived for the high amplitude variable V7 (0.779); we conclude that
the radial modes excited in V7 are the fundamental mode and first
overtone.  The period ratios derived for V2 and V8 (0.781 and 0.780
respectively) are also very similar. We have already assumed that the
longer of the periods detected in V8 corresponds to the radial mode (on
the basis of its high amplitude), and it seems likely that this is the
fundamental mode and that the other mode is the first overtone.  In the
case of V2 both modes are of low amplitude, and the period ratio is the
only argument in favour of the first overtone/fundamental mode
interpretation. However this conclusion may not be correct.  Poretti
(2000) gives the example of the $\delta$ Scuti star 44 Tau to show that
nonradial modes can also show a similar period ratio. Several examples
of period ratios in the range 0.76-0.78 can be found among the seven
terms observed in this star.

In the case of low amplitude variable V5, the situation is even more
complicated. Three frequencies have been detected in the Fourier
spectrum, and some additional periodicities seem to be visible in the
residuals.  We are not able to identify frequencies unambiguously
because of daily aliases.  When dividing some of the possible
frequencies we get period ratios that fall in the range 0.77-0.79
typical for $P_1/P_0$ (22.991/29.367 = 0.783), but values typical for
$P_2/P_1$ (22.991/28.204 = 0.815) are also possible. From the position
on the period-luminosity diagram (Fig. 14) the interpretation of 22.991
as the first overtone seems more likely, but we cannot exclude the
possibility that it is a nonradial mode. Further observations are
needed to analyze the frequency spectrum of this star.

In the case of the other stars (V1,V3,V4,V10) we observed two closely
spaced periodicities - at least one of these must correspond to a
nonradial mode.  The same applies to low amplitude modes present in the
vicinity of the high amplitude ones in V6, V9 and V11.

\section{Period-luminosity relation and distance modulus to NGC~3201}
In Fig. 14 we show the period-luminosity relation for the SX~Phoenicis
stars in NGC~3201. Circles represent positions of the presumed radial
modes (assigned on the basis of the full visual amplitude $A_V > 0.1$
mag).  The horizontal lines connect the pairs of modes in variables V2,
V7 and V8, for which the observed period ratio suggests pulsations in
fundamental mode and first overtone.  Also plotted are lines
corresponding to McNamara's (1997) period-luminosity relations for the
fundamental mode ($M_V=-3.725 \log P_0-1.933$) and first overtone,
shifted by the distance modulus to the cluster. We adopted a
period-ratio of  $P_1/P_0=0.779$ when  calculating the first overtone
relation.

\begin{figure}
\epsfxsize=8cm\epsffile{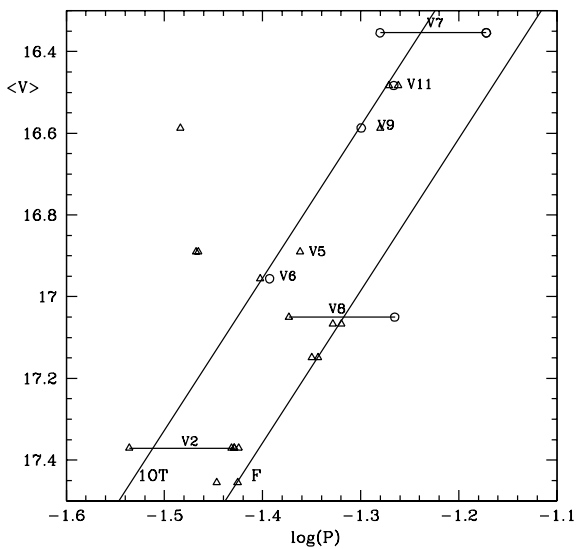}
\caption{The P-L relation of SX~Phoenicis stars in NGC~3201 (see text for details).
}
\label{fig2}
\end{figure}

The apparent distance modulus to the cluster has been calculated as a
mean of distant moduli determined from McNamara's (1997) P-L relation
for 6 variables (V2, V6, V7, V8, V9, V11). The individual distance
moduli were 13.98, 14.11, 13.91, 14.27, 14.09, 14.10, respectively. The
variables V6, V9 and V11 have been assumed to be pulsating in the first
overtone and corresponding fundamental periods were calculated assuming
a period ratio $P_1/P_0=0.779$.  We obtain a mean apparent distance
modulus to NGC~3201 $(m-M)_V=14.08\pm 0.06$. For comparison
a mean distance modulus calculated taking into account only 3 variables
pulsating in the first overtone
and fundamental mode is $(14.05\pm 0.11$).
The quoted error does not
include systematic errors in McNamara's period-luminosity relation.  To
estimate the systematic error we calculated the distance modulus to
NGC~3201 using two other P-L relations. Pych et al. (2001) presented
P-L relations for SX~Phe stars in the globular cluster M55. These
fundamental mode and first overtone relations yield $(m-M)_V=13.99$ and
$(m-M)_V=13.97$, respectively.  The Petersen \& H\o g 1998 relation  is
based on {\it Hipparcos} parallaxes and gives a mean modulus of
$(m-M)_V=14.02$. We estimate that the systematic error in the
determination of distance moduli based on P-L relations is $\sim$0.1
mag, so the final result for the distance modulus to NGC~3201 is
$(m-M)_V=14.08\pm 0.06\pm 0.1$.

\setcounter{table}{3}
\begin{table*}
\begin{minipage}{160mm}
\caption{Recent distance modulus determinations to NGC~3201}
\label{tab4}
\begin{tabular}{@{}c c c l l }
$(m-M)_V$&$(m-M)_0$ & $<E(B-V)>$%
\footnote{NGC~3201 is affected by differential reddening, in column (3) we give a mean value for the cluster }& method       &  reference       \\
       &        &         &              &                  \\      
 14.24 & 13.58  & 0.21%
 \footnote{average of different sources}   &$ M_V(HB)=0.15[Fe/H]+0.8$    &  Harris 1996     \\
 13.95 & 13.27       & 0.22%
 \footnote{adopted} & local subdwarfs fitting & Covino \& Ortolani 1997\\
 14.15 & 13.47   & 0.22{\phantom{$^a$}} & subdwarf fiducial fitting& Covino \& Ortolani 1997\\
 14.0{\phantom 0}& 13.32       & 0.22{\phantom{$^a$}}  & RR~Lyrae          & Covino \& Ortolani 1997 \\
 14.2{\phantom 0}& 13.52       & 0.22{\phantom{$^a$}} & RR~Lyrae          & Covino \& Ortolani 1997 \\
 13.84 & 13.26  & 0.17{\phantom{$^a$}}  & isochrone fitting & von Braun \& Mateo 2001\\
 14.25 & 13.32  & 0.30{\phantom{$^a$}} & RR~Lyrae          & Piersimoni et al. 2002 \\
 14.21 & 13.35  & 0.28{\phantom{$^a$}}  & RR~Lyrae          & Layden 2002            \\
 14.08 &        &         & SX~Phe            & this paper             \\

\end{tabular}
\end{minipage}
\end{table*}

Our value of the distance modulus to NGC 3201 is only slightly lower
than the distance modulus listed in the Harris (1996) Catalog of
Parameters for Globular Clusters in the Milky Way (the electronic
edition as updated in June 1999) ($(m-M)_V=14.24$). The distance scale
adopted there is defined by $M_V(HB)=0.15[Fe/H]+0.8$.  Covino \&
Ortolani (1997) measured the distance modulus to NGC~3201 by main
sequence fitting to the subdwarf sequence and by the $M_V-[Fe/H]$
relation for RR~Lyrae stars. Main sequence fitting to 8 local subdwarfs
yielded $(m-M)_V=13.95$, and the use of the subdwarf fiducial resulted
in $(m-M)_V=14.15$.  However the absolute magnitudes of the field
subdwarfs they used had been derived from ground based parallaxes which
are known now to have the systematic errors and which are larger than
those determined by {\it Hipparcos}.  The RR~Lyraes yielded two values
$(m-M)_V=14.0$ and $14.2$, depending on the $M_V-[Fe/H]$ relation
used.  von Braun \& Mateo (2001) obtained a true distance modulus
$(m-M)_0=13.26$ and an average reddening $<E(V-I)>=0.24$ from
isochrone fitting, resulting in an observed distance modulus
$(m-M)_V=13.84$.  Piersimoni, Bono, \& Ripepi (2002) determined the
distance to NGC 3201  based on cluster RR~Lyrae stars. They measure
($(m-M)_0=13.32$ and $<E(B-V)>=0.3$) or $(m-M)_V=14.25$ for $R_V=3.1$.
Layden (2002) also used observations of RR Lyrae stars to obtain a true
distance modulus $(m-M)_0=13.35$ and mean $<E(B-V)>=0.28$, or
$(m-M)_V=14.21$.
Our distance modulus to NGC~3201, $(m-M)_V=14.08\pm
0.06\pm 0.1$, falls in the range defined by the other determinations.
Recent distance modulus determinations are summarized in Table 4.
\section{Summary}
We have surveyed the southern globular cluster NGC~3201  for
short-period variable stars. In this first of a series of papers we
present  CCD photometry of eleven multimode SX~Phoenicis--type blue
stragglers discovered in this cluster. Fourier periodograms of the $V$-band 
light curves are presented. The analysis of Fourier transforms
shows that both radial and non-radial modes are excited. A tentative
identification of radial modes in six variables is proposed.  For
three  stars (V2, V7, V8) the derived period ratios are close to that
observed in SX~Phoenicis itself,  suggesting pulsations in the
fundamental and the first-overtone radial modes.  For three more stars
(V6, V9 and V11) a radial mode has been assigned on the basis of high
($>$ 0.3 mag) or medium ($>$ 0.1 mag) amplitudes of variability; from
the position on period-luminosity diagram we infer that in these
variables the first overtone is excited.  In eight of the variables
pairs of closely spaced modes are detected that cannot be explained
with radial modes, at least one of them must be nonradial.  The
apparent distance modulus to the cluster has been calculated from
distant moduli determined from McNamara's (1997) P-L relation applied
to six presumed radial pulsators. We measure
$(m-M)_V=14.08\pm0.06\pm0.1$, consistent with other determinations.

\section*{Acknowledgments}

BM and WK were supported by the grant 5.P03D.004.21 from the Committee
of Scientific Research (Poland).  IBT and WK were supported by NSF
grant AST-9819786.  We would like to thank Dr. Paul Schechter for
making {\sc d{\footnotesize o}phot} available to us, and Dr. Alex
Schwarzenberg-Czerny for the opportunity to use his programme {\sc
NFIT}. 
We also thank the director of The Observatories of the Carnegie
Institution of Washington for a generous allocation of observing time.
BM acknowledges with gratitude the kind hospitality of the Las Campanas
Observatory during her stays in Chile, and WK the generous hospitality
of the Copernicus Astronomical Centre where this paper was written.
An anonymous referee's comments are also appreciated.
This research has made use of NASA's Astrophysics Data System Abstract
Service.

\end{document}